\documentclass[10pt,letterpaper]{article}
\usepackage[top=0.85in,left=2.75in,footskip=0.75in]{geometry}

\usepackage{amsmath,amssymb}

\usepackage{changepage}

\usepackage{textcomp,marvosym}

\usepackage{cite}

\usepackage{nameref,hyperref}

\usepackage[right]{lineno}

\usepackage[nopatch=eqnum]{microtype}
\DisableLigatures[f]{encoding = *, family = * }

\usepackage[table]{xcolor}

\usepackage{array}

\newcolumntype{+}{!{\vrule width 2pt}}

\newlength\savedwidth

\raggedright
\usepackage[aboveskip=1pt,labelfont=bf,labelsep=period,justification=raggedright,singlelinecheck=off]{caption}

\makeatletter
\renewcommand{\@biblabel}[1]{\quad#1.}
\makeatother

\usepackage{lastpage,fancyhdr,graphicx}
\usepackage{epstopdf}
\fancyheadoffset[L]{2.25in}
\fancyfootoffset[L]{2.25in}
\begin{document}
\vspace*{0.2in}

\begin{flushleft}
{\Large
\textbf{GTIN: A Scalable Unified Framework for Joint Event and Time Prediction in Temporal Graphs} % Please use "sentence case" for title and headings (capitalize only the first word in a title (or heading), the first word in a subtitle (or subheading), and any proper nouns).
}
\newline
\\
Mohammad Ostadmohammadi\textsuperscript{1},
Sepehr Kazemi\textsuperscript{2},
Hamid R. Rabiee\textsuperscript{1*}
\\
\bigskip
\textbf{1} Department of Computer Engineering, Sharif University of Technology, Tehran, Iran
\\
\textbf{2} Department of Electrical Engineering, Sharif University of Technology, Tehran, Iran
\\
\bigskip

% Insert additional author notes using the symbols described below. Insert symbol callouts after author names as necessary.
% 
% Remove or comment out the symbols in the byline and the author notes below if they aren't used.
%

% Use the asterisk to denote corresponding authorship and provide email address in note below.
* rabiee@sharif.edu

\end{flushleft}

% For PLOS Medicine research article authors, please structure your abstract
% with "Background", "Method and Findings" and "Conclusion" sections per
% journal requirements.

% For PLOS Neglected Tropical Diseases research article authors, please
% structure your abstract with "Background", "Methodology", "Findings", and
% "Conclusion" sections per journal requirements.
%
\section*{Abstract}
Temporal graphs are increasingly used to model dynamic systems across diverse domains, including social, financial, and traffic networks. Predicting both what the next event will be and when it will occur in these systems is crucial for understanding and anticipating complex behaviors, but has not been studied much yet, especially in large-scale graphs.

To address this gap, we propose a scalable unified mathematical framework that captures varying degrees of complexity across temporal graphs with events occurring on the edges. Our framework is flexible and expressive enough to accommodate a wide range of network structures and temporal dynamics. Building upon this analysis, we introduce our novel scalable approach for jointly predicting the next event and its occurrence time. Empirical evaluations across multiple datasets demonstrate that our method consistently outperforms existing techniques, particularly in scenarios involving irregular event patterns and complex temporal dependencies. These findings highlight the potential of our framework as a robust foundation for future research in temporal event prediction.

\newgeometry{top=0.85in,left=1in,right=1in,footskip=0.75in}

% Use "Eq" instead of "Equation" for equation citations.
\section*{Introduction}
\subsection*{Problem Statement}
The prediction of future events in temporal graphs has emerged as a central problem in modeling complex temporal systems. Applications such as communication networks, social media, transportation systems, and biological interactions depend on the ability to anticipate not only what event will occur next, but also \textit{when} and \textit{where} it will happen. Traditional static graph models fail to capture temporal dependencies, while existing temporal graph approaches often focus on localized subgraphs or are restricted in scalability.

Recent developments in temporal graph neural networks (TGNNs) and temporal point process (TPP) models have made progress in understanding the temporal evolution of graph structures. However, most existing models operate within limited temporal windows or node neighborhoods, failing to utilize the global structural context. Furthermore, while several methods address link or event prediction, only a few explicitly model both the event occurrence time and its location or lack scalability to extend to large graphs.

\subsection*{Aim and Achievements}
To address these limitations, we introduce a \textbf{Global Temporal Interaction Network (GTIN)}, a unified framework capable of predicting the next event's \textit{time} and \textit{location} by considering the graph as a whole. The model integrates hierarchical decomposition and temporal aggregation, enabling scalable training and inference even on large-scale temporal graphs.

The main contributions of this work can be summarized as follows:

\begin{itemize}
    \item \textbf{Global Event-Time Prediction Framework:} We propose a model that jointly predicts the next event's location and timing by analyzing the entire graph structure rather than relying solely on local neighborhoods.
    \item \textbf{Hierarchical Scalability:} The proposed model introduces a hierarchical graph decomposition mechanism that allows any graph, regardless of size, to be efficiently partitioned and processed while maintaining temporal dependencies.
\end{itemize}

To validate the effectiveness and scalability of the proposed GTIN model, we conduct extensive experiments on multiple benchmark temporal graph datasets. The evaluation focuses on \textit{time estimation precision}, comparing our approach against leading temporal graph baselines such as TGAT, TGN, and DyRep. We further assess scalability through hierarchical partitioning experiments on large graphs, demonstrating the model’s efficiency and robustness under diverse dynamic environments.

\subsection*{Related Works}
Research on event prediction in temporal graphs can be broadly categorized into four main directions: (1) Temporal Point Process Models, (2) Temporal Graph Neural Networks, (3) Hybrid Temporal-Spatial Event-Time Models, and (4) Scalable and Hierarchical Graph Learning Frameworks. This section reviews representative works in each category and highlights their connections and limitations with respect to our study.

\subsubsection*{Temporal Point Process Models}
Temporal Point Processes (TPPs) provide a classical foundation for modeling event occurrences over continuous time. The Hawkes process \cite{hawkes1971spectra}, a self-exciting TPP, captures dependencies between past and future events through intensity functions. Neural extensions, such as the Neural Hawkes Process~\cite{mei2017neural}, learn complex temporal patterns using neural representations. Although these methods accurately model event timing, they generally ignore the underlying graph structure, limiting their applicability to relational domains.

Recent advances in temporal point processes have improved joint event and time prediction. Integration-free Marked TPP \cite{liu2026addressing} addresses event-mark imbalance, while Continuous-Time State-Space Models \cite{chang2026deep} and Mixed-Type Event Transformers \cite{draxler2026transformers} provide efficient architectures for modeling complex event sequences. However, these approaches do not explicitly incorporate global graph structures.

\subsubsection*{Temporal Graph Neural Networks}
Temporal Graph Neural Networks (TGNNs) extend static GNNs to dynamic settings, enabling the modeling of both structural and temporal dependencies. Models such as TGAT~\cite{xu2020inductive}, TGN~\cite{rossi2020temporal}, and DyRep~\cite{trivedi2019dyrep} leverage temporal message passing and node memory mechanisms to track evolving node states. More recent developments, including MTGN~\cite{liu2023mtgn}, HierTKG~\cite{almutairi2024hierpkg}, and RE-GAT~\cite{li2023future}, have focused on temporal knowledge graphs, predicting future relations or interactions between entities. However, these models typically rely on local temporal neighborhoods and thus struggle to capture the \textit{global} graph context required for comprehensive event forecasting.

Recent approaches increasingly incorporate structural and temporal dependencies into dynamic graph learning. Global-Local Temporal Graph Learning \cite{wang2025boosting} captures structural information at multiple scales, while DNRHP \cite{ying2025dnrhp} and Time-Aware Graph Learning \cite{pan2025time} model temporal interactions for network representation learning and link prediction. These methods do not explicitly address graph-wide joint event and time prediction.

\subsubsection*{Hybrid Temporal-Spatial and Event-Time Prediction Models}
Hybrid frameworks combine spatial, temporal, and event-level reasoning. For example, GSTPP~\cite{zhou2025fine} employs a graph spatio-temporal point process with a self-adaptive anchor graph to predict fine-grained event occurrences in both space and time. MTGN~\cite{liu2023mtgn} jointly models missing and observed events to improve prediction robustness under partial observability. Knowledge-enhanced approaches, such as CEEG~\cite{wang2025ceeg} and TPAR~\cite{chen2024tpar}, incorporate commonsense reasoning or interpolation-extrapolation mechanisms to improve long-term temporal predictions. Despite these advances, such models often suffer from high computational complexity and lack scalability to large graphs.

Recent hybrid frameworks further integrate structural and temporal information. EventFormer \cite{yu2025eventformer} employs hierarchical neural point processes for spatio-temporal event prediction, while DGNPP \cite{chen2025deep} combines graph neural networks with temporal point processes to predict future interactions and their occurrence times. In contrast, GTIN employs a global graph representation for next-event location and time prediction.

\subsubsection*{Scalable and Hierarchical Graph Learning Frameworks}
Scalability has become a key challenge in temporal graph learning. Approaches like TF-TGN~\cite{xu2023tftgn}, SIMPLE~\cite{rossi2023simple}, and TAP-GNN~\cite{fang2023tapgnn} improve efficiency through temporal sampling, memory compression, and hierarchical training. While these models enhance scalability, they often operate on fixed-size graph partitions and fail to exploit the network's global temporal structure. In contrast, our proposed method introduces a \textbf{hierarchical decomposition and aggregation} strategy that maintains both temporal and structural dependencies across scales, enabling global-level event-time prediction while preserving computational efficiency.

Recent frameworks have improved the efficiency of temporal graph learning. UTG \cite{huang2024utg} unifies snapshot-based and event-based temporal graph modeling, while ScaDyG \cite{wu2026scadyg} addresses large-scale dynamic graph learning. Unlike these methods, GTIN combines DGI-based representation learning and balanced clustering for scalable hierarchical event-time prediction.

\subsubsection*{Summary and Positioning}
Recent advances in temporal graph learning and temporal point process modeling have significantly improved the ability to capture complex structural and temporal dependencies in dynamic networks. Existing approaches have explored various strategies for modeling event sequences, incorporating local and global graph information, jointly predicting events and their occurrence times, and improving computational efficiency through scalable architectures. Despite these developments, several challenges remain in simultaneously capturing graph-wide temporal dependencies, accurately predicting both the location and timing of the next event, and maintaining computational efficiency in large-scale temporal graphs. To address these challenges, our proposed Global Temporal Interaction Network (GTIN) integrates temporal message passing with a global graph representation to jointly estimate the next event's location and occurrence time. Furthermore, by incorporating hierarchical graph decomposition through unsupervised representation learning and balanced clustering, GTIN improves scalability while preserving predictive performance. Consequently, our framework provides a unified approach that combines global temporal reasoning, joint event-time prediction, and hierarchical processing within a single architecture.

\section*{Materials and methods}
\subsection*{Preliminaries}
\subsubsection*{Simple Point Process}

Consider the sequence, \( t_1, t_2, \ldots, t_n \). We define \( f(t|\mathcal{H}_{t_n}) \) as a conditional density function.  
The conditional intensity function for \( t > t_n \),
\begin{eqnarray}
\lambda(t) = \frac{f(t|\mathcal{H}_{t_n})}{1 - F(t|\mathcal{H}_{t_n})}
\end{eqnarray}
is used more often than \( f(t|\mathcal{H}_{t_n}) \). We can say that \( \lambda(t) = f(t|\mathcal{H}_{t^-}) \). From the definition, we have
\begin{eqnarray}
f(t|\mathcal{H}_{t_n}) = \lambda(t) \exp \left( - \int_{t_n}^{t} \lambda(\tau) d\tau \right)
\end{eqnarray}

The Log Likelihood of an observed sequence \( t_1, t_2, \ldots, t_n \) in an interval \([0, T]\) is defined as:

\begin{eqnarray}
LL = \log[f(t_1, t_2, \ldots, t_n)(1 - F(T | \mathcal{H}_{t_n}))]
\end{eqnarray}

$$
= \log \left( \prod_{i=1}^n f(t_i | \mathcal{H}_{t_{i-1}})(1 - F(T | \mathcal{H}_{t_n})) \right)
$$

$$
= \left( \sum_{i=1}^n log(\lambda(t_i)) - \int_{t_{i-1}}^{t_i} \lambda(\tau) \, d\tau \right) - \int_{t_n}^T \lambda(\tau) \, d\tau
$$

$$
= \sum_{i=1}^n log(\lambda(t_i)) - \int_0^T \lambda(\tau) \, d\tau
$$

\subsubsection*{Marked Point Process}

Consider the sequence, \((t_1, k_1), (t_2, k_2), \ldots, (t_n, k_n)\). We define \(p(t, k|\mathcal{H}_{t^-}) = p(t|\mathcal{H}_{t^-})p(k|t, \mathcal{H}_{t^-})\), 
where \(p\) stands for both PMF and PDF. The conditional intensity function for \(t > t_n\) is defined as:
\begin{eqnarray}
\lambda(t, k) = p(t|\mathcal{H}_{t^-})p(k|t, \mathcal{H}_{t^-})
\end{eqnarray}
$$
= \lambda(t)p(k|t, \mathcal{H}_{t_n}) = \frac{p(t|\mathcal{H}_{t_n})p(k|t, \mathcal{H}_{t_n})}{1 - F(t|\mathcal{H}_{t_n})}
$$
$$
= \frac{p(t, k|\mathcal{H}_{t_n})}{1 - F(t|\mathcal{H}_{t_n})}
$$
Also, we have:
\begin{eqnarray}
p(t, k|\mathcal{H}_{t_n}) = \lambda(t, k) \exp \left(- \int_{t_n}^{t} \lambda(\tau) d\tau \right)
\end{eqnarray}
Now, we derive the log-likelihood function for the marked sequence \((t_1, k_1), (t_2, k_2), \ldots, (t_n, k_n)\) in an interval \([0, T]\):
\begin{eqnarray}
LL = \log \left[ p(t_1, k_1, t_2, k_2, \ldots, t_n, k_n)(1 - F(T|\mathcal{H}_{t_n})) \right]
\end{eqnarray}
$$
= \sum_{i=1}^{n} log(\lambda(t_i, k_i)) - \int_{0}^{T} \lambda(\tau) d\tau
$$

\subsubsection*{Problem Formulation}

We have a graph with fixed nodes in which events occur on the edges. Our objective is to predict where and when the next event occurs. We define \( \mathcal{H}_n(t) = \left\{ t_i^{(n)} \;|\; t_i^{(n)} \leq t \right\} \), i.e., the history of the \( n\)th edge until time \( t \),  
where \( t_i^{(n)} \) is the time of the \( i\)th interaction.

We define 
\begin{eqnarray}
\mathcal{H}_n(t, L) = \left\{ t_i^{(n)} \;|\; t \leq t_i^{(n)}, 1 \leq i \leq L \right\}
\end{eqnarray}
i.e., the next \( L \) interactions. 

Also we define 
\begin{eqnarray}
\mathcal{H}(t) = \left\{ t_i^{(n)} \mid t_i^{(n)} \leq t, 1 \leq n \leq N \right\}
\end{eqnarray}
i.e., the history of the graph until time \( t \), and 
\begin{eqnarray}
\mathcal{H}(t, L) = \left\{ t_i^{(n)} \mid t \leq t_i^{(n)}, 1 \leq i \leq L, 1 \leq n \leq N \right\}
\end{eqnarray}
i.e., the next \( L \) interactions of the graph after time \( t \).

\textbf{Definition (Problem):} Assume until time \( t \), we observe \( \mathcal{H}(t) \). Our objective is to estimate \( \mathcal{H}(t, 1) \).

You may wonder why we didn’t define our problem as predicting \( \mathcal{H}(t, L) \). Actually, if we can predict \( \mathcal{H}(t, 1) \), 
we can repeat this process \( L \) times. The model in \cite{xia2022graph} considers the problem as predicting the next event time of each edge based on its history and its neighbors, 
not considering the whole graph structure. But in real-world applications, we need to estimate events in an interval considering all edges. 
In other words, in an interval, an edge may have many interactions, but some edges have no interactions. Also, the interaction on some edges 
changes the history of the graph, which can affect the prediction of other edges.

\subsubsection*{Objective}

To estimate the \( \mathcal{H}(t, 1) \) as well as possible, we should maximize the following objective:
\begin{eqnarray}
P(\mathcal{H}(t, 1), \mathcal{H}(t)) = P(\mathcal{H}(t)) P(\mathcal{H}(t, 1) | \mathcal{H}(t))
\end{eqnarray}

For the first term, we should compute \( P(\mathcal{H}(t)) \). Without loss of generality, we write
\( \mathcal{H}(t) = \{(t_i, c_i) | t_i \leq t \}_{i=1}^{K} \), where \( K \) is the total number of events until time \( t \) and
\( c_i \) is an integer in range \([1, N]\) showing the edge that this \( t_i \) belongs to. Also, \( t_i \)'s are ordered.
Now we can write the first term as:

\begin{eqnarray}
P(\mathcal{H}(t)) = p(t_1, c_1, t_2, c_2, \ldots, t_{K}, c_{K})
\end{eqnarray}

$$
= \prod_{i=1}^{K} p(t_i, c_i | \mathcal{H}(t_{i-1}))
$$

$$
= \prod_{i=1}^{K} p(c_i | \mathcal{H}(t_{i-1})) p(t_i | c_i, \mathcal{H}(t_{i-1}))
$$

This is the likelihood of a marked point process where the marks are \(  c_i \)'s. 
Now, the probability of this likelihood can also be written as:

\begin{eqnarray}
P(\mathcal{H}(t)) = \prod_{i=1}^{K} \lambda(t_i, c_i) \exp \left( - \int_{0}^{t} \lambda(\tau) d\tau \right)
\end{eqnarray}

Which format we write depends on our implementation, but working with density functions are easier because estimating the integral in intensity functions can lead to poor performance.

Now we consider the second term, \( P(\mathcal{H}(t, 1) | \mathcal{H}(t)) \), where we want to predict the next interaction on the graph. So we have:
\begin{eqnarray}
P(\mathcal{H}(t, 1) | \mathcal{H}(t)) = p(\hat{t}, \hat{c} | \mathcal{H}(t)) = p(\hat{c} | \mathcal{H}(t)) p(\hat{t} | \hat{c}, \mathcal{H}(t))
\end{eqnarray}

So we can write the final objective as:
\begin{eqnarray}
\text{maximize} \quad \sum_{i=1}^{K} \log p(c_i | \mathcal{H}(t_{i-1})) + \log p(t_i | c_i, \mathcal{H}(t_{i-1})) + \log p(\hat{c} | \mathcal{H}(t)) + \log p(\hat{t} | \hat{c}, \mathcal{H}(t))
\end{eqnarray}

Or in terms of intensity function:
\begin{eqnarray}
\text{maximize} \quad \sum_{i=1}^{K} \log \lambda(t_i, c_i) - \int_0^t \lambda(\tau) d\tau + \log p(\hat{c} | \mathcal{H}(t)) + \log p(\hat{t} | \hat{c}, \mathcal{H}(t))
\end{eqnarray}

\subsubsection*{Model}

In this section, we introduce our implementation of the model. As mentioned previously, the computation of the integral of the conditional intensity function in the likelihood can require numerical approximation and may reduce the stability of the model. Therefore, instead of directly estimating the intensity function, our goal is to estimate the following conditional density function:

\begin{eqnarray}
p(t_i, c_i | \mathcal{H}(t_{i-1})) = p(c_i | \mathcal{H}(t_{i-1}))p(t_i | c_i, \mathcal{H}(t_{i-1}))
\end{eqnarray}

This decomposition follows the natural order of the prediction problem. First, the model predicts the location of the next event, which is represented by $c_i$. Then, conditioned on the predicted edge and the previous history, the model predicts the occurrence time $t_i$. Therefore, the model jointly answers two questions: where the next event happens and when it happens.

Now we take care of $\mathcal{H}(t_i)$. For computing this, we consider that each edge has an embedding $h_n(t)$ up to some time $t$. Each edge in the original graph is treated as a node in the transformed graph, and neighboring edge-nodes are those that share a common endpoint in the original graph. So we use the Message Passing Framework as follows:

\begin{eqnarray}
h_n(t) =
\text{COMB}\{
\text{AGG}
\{
\phi(t_i^j) | t_i^j \leq t, j \in \mathcal{N}(n)
\},
\phi(t_i^n) | t_i^n \leq t
\}
\end{eqnarray}

, where COMB uses self-attention mechanism and AGG a simple learnable linear transformation.

In this equation, $\mathcal{N}(n)$ is the set of neighbors of edge $n$. The first part aggregates the temporal information coming from neighboring edges, while the second part keeps the temporal information of the edge itself. Therefore, the embedding $h_n(t)$ is not only based on the past events of edge $n$, but also on the past events of its neighboring edges. One can stack multiple layers to consider more hops and to propagate temporal information to a larger neighborhood of the graph.

The function $\Phi(\cdot)$ converts each temporal interaction into a vector representation. Since time is a continuous variable, using the raw timestamp alone is not enough to represent complex temporal patterns. Therefore, following the time conversion strategy used in GNPP \cite{xia2022graph}, we encode each timestamp using a harmonic time encoder:

\begin{eqnarray}
\phi_h(t) =
\sqrt{\frac{1}{d_\omega}}
[
\cos(\omega_1 t),
\sin(\omega_1 t),
\ldots,
\cos(\omega_{d_\omega} t),
\sin(\omega_{d_\omega} t)
]
\end{eqnarray}

where $\omega_1, \omega_2, \ldots, \omega_{d_\omega}$ are learnable frequency parameters. This transformation maps each continuous timestamp into a high-dimensional vector. The advantage of this conversion is that the model can capture different temporal patterns, such as short-term recency effects, long-term dependencies, and possible periodic behavior, without discretizing the temporal graph into fixed snapshots.

Finally, the embedding of the graph up to time $t$ is the sum of all of its edge embeddings:

\begin{eqnarray}
h_G(t) = \sum_{n=1}^{N} w_n h_n(t)
\end{eqnarray}

where $N$ is the number of edges in the graph and $w_n$ is the weight of edge $n$. This weight can be fixed, learned, or computed by an attention mechanism. The graph embedding $h_G(t)$ is used as the representation of the whole history $\mathcal{H}(t)$. This is important because the next event is not only affected by the history of one edge, but also by the global temporal state of the whole graph.

Now we explain the two terms of the likelihood in more detail. The first term $p(c_i | \mathcal{H}(t_{i-1}))$ is a classification problem, because $c_i$ is a discrete variable showing the edge on which the next event occurs. Therefore, we use a softmax function over all candidate edges:

\begin{eqnarray}
p(c_i=n | \mathcal{H}(t_{i-1})) =
\text{softmax}
(
W_c h_G(t_{i-1}) + b_c
)_n
\end{eqnarray}

This term gives the probability that edge $n$ will be the location of the next event. It forces the model to learn which parts of the graph are more likely to become active based on the previous temporal history.

The second term $p(t_i | c_i, \mathcal{H}(t_{i-1}))$ is a regression problem. Since the occurrence time is continuous, we consider a Gaussian distribution for this term. Therefore, the model has to predict both the mean and the variance of the time distribution. For this purpose, we use both the global graph embedding and the embedding of the selected edge:

\begin{eqnarray}
z_i^t =
[
h_G(t_{i-1}) \Vert h_{c_i}(t_{i-1})
]
\end{eqnarray}

Then, the parameters of the time distribution are obtained by a neural network:

\begin{eqnarray}
\mu_i^t, s_i^t =
\text{MLP}_t(z_i^t)
\end{eqnarray}

\begin{eqnarray}
\sigma_i^t =
\text{softplus}(s_i^t) + \epsilon
\end{eqnarray}

where $\epsilon$ is a small positive value added for numerical stability. Therefore, the time density is written as:

\begin{eqnarray}
p(t_i | c_i, \mathcal{H}(t_{i-1})) =
\mathcal{N}
(
t_i ;
\mu_i^t,
(\sigma_i^t)^2
)
\end{eqnarray}

In implementation, this distribution can be applied either to normalized timestamps or to normalized inter-event times. The mean $\mu_i^t$ represents the expected occurrence time of the next event, while the variance $(\sigma_i^t)^2$ represents the uncertainty of this prediction. Thus, the model is not only penalized for predicting an inaccurate time, but also for assigning an unsuitable uncertainty to the prediction.

Based on the two terms above, the negative log-likelihood loss for a sequence of $K$ events is:

\begin{eqnarray}
-\sum_{i=1}^{K}
[
\log p(c_i | \mathcal{H}(t_{i-1}))
+
\log p(t_i | c_i, \mathcal{H}(t_{i-1}))
+
\log p(W_i | t_i, c_i, \mathcal{H}(t_{i-1}))
]
\end{eqnarray}

This loss can also be interpreted as the sum of two different parts:

\begin{eqnarray}
\mathcal{L}{c}
+
\lambda_t \mathcal{L}{t}
\end{eqnarray}

where $\mathcal{L}{c}$ is the classification loss for predicting the next edge and $\mathcal{L}{t}$ is the time prediction loss. The parameter $\lambda_t$ controls the importance of the time. This decomposition makes the training objective more interpretable, because each part corresponds to one aspect of the next-event prediction problem.

In training, we use a teacher forcing approach. At each step $i$, the model receives the true history $\mathcal{H}(t_{i-1})$, and the likelihood is computed using the true event tuple $(t_i, c_i)$. After the loss is computed, the true event is added to the history, and the model moves to the next step. This helps the model learn from correct historical contexts and prevents error accumulation during training.

In inference, the model receives the observed history $\mathcal{H}(t)$ and first computes $h_G(t)$ and the edge embeddings $h_n(t)$. Then, it computes $p(c | \mathcal{H}(t))$ for candidate edges. For each candidate edge $c$, the model predicts the mean of the time distribution:

\begin{eqnarray}
\hat{t}_c = \mu_c^t
\end{eqnarray}

Finally, the model selects the tuple with the maximum joint probability:

\begin{eqnarray}
\arg\max_{t,c}
[
\log p(c | \mathcal{H}(t))
+
\log p(t | c, \mathcal{H}(t))
]
\end{eqnarray}

% Results and Discussion can be combined.
\section*{Results}
\subsection*{Baseline methods}
In this section, we discuss baselines we used in experiments, including three neural point process models, i.e., Transformer Hawkes Process (THP) \cite{zuo2020transformer}, Self-Attentive Hawkes Process (SAHP) \cite{zhang2020self}, and Neural Network Point Process (NNPP) \cite{omi2019fully}, and seven graph neural models, i.e., TGAT \cite{xu2020inductive}, TGN \cite{rossi2020temporal}, GHNN \cite{han2020graph}, HTNE \cite{zuo2018embedding}, DySAT \cite{sankar2020dysat}, CTDNE \cite{nguyen2018continuous}, and GAT-LSTM. GAT-LSTM is a GAT model equipped with the harmonic encoder and an LSTM module as a naive GNN extended baseline, while others are originally devised for temporal graphs and GNPP model \cite{xia2022graph}.

\subsection*{Datasets}
\subsubsection*{Real-World datasets}

Wikipedia is a temporal graph containing edited pages and users as nodes. Each edition between a user and an page is recorded with a timestamp. This temporal graph contains about 9300 nodes and 160000 temporal edges. \cite{kumar2019predicting}.

Reddit is an online community that records interactions between active users and their posts under subreddits. It has about 11000 nodes and 700000 temporal edges. \cite{kumar2019predicting}.

CollegeMsg is a communication graph that records private Email interactions of an online society at the University of California, Irvine. This temporal graph contains about 2000 users and 60000 temporal edges \cite{panzarasa2009patterns}.

\subsubsection*{Synthetic datasets}
These datasets are generated as mentioned in \cite{xia2022graph}.In this paper, firstly a random Watts–Strogatz small-world graph is generated with 500 nodes and 1000 edges. Secondly, using Temporal Point Processes edge events are generated by Hawkes-negative, Hawkes-positive and Poisson process.  

\subsection*{Evaluation Metrics}

We evaluate GTIN from
two complementary perspectives: time prediction accuracy and next-event
location prediction.

Let $\mathcal{H}(t_{i-1})$ denote the observed graph history before the $i$th
test event, let $c_i$ denote the edge on which the event occurs, and let $t_i$
denote its occurrence time. All predictions are made chronologically using only
events observed before $t_i$.

For time prediction, we retain the root mean squared error:
\begin{eqnarray}
\operatorname{RMSE}
=
\sqrt{
\frac{1}{K}
\sum_{i=1}^{K}
\left(
t_i-\hat{t}_i
\right)^2
},
\label{eq:rmse}
\end{eqnarray}
where $K$ is the number of test events and $\hat{t}_i$ is the predicted
occurrence time.

For location prediction, the model assigns a probability to every candidate
edge:
\begin{eqnarray}
p\left(c_i=n \mid \mathcal{H}(t_{i-1})\right),
\qquad n\in\mathcal{C}_i,
\label{eq:location_probability}
\end{eqnarray}
where $\mathcal{C}_i$ is the candidate edge set for event $i$. The rank of the
true edge is denoted by $r_i$. We report Hits@$q$ and mean reciprocal rank
(MRR), defined as
\begin{eqnarray}
\operatorname{Hits@}q
=
\frac{1}{K}
\sum_{i=1}^{K}
\mathbb{I}\left[r_i\leq q\right],
\label{eq:hits}
\end{eqnarray}
and
\begin{eqnarray}
\operatorname{MRR}
=
\frac{1}{K}
\sum_{i=1}^{K}
\frac{1}{r_i}.
\label{eq:mrr}
\end{eqnarray}
Higher Hits@$q$ and MRR values indicate better next-event location predictions.

\subsection*{Results on real-world datasets}
As shown in tab~\ref{tab:table1}, our Global Temporal Interaction Network (GTIN) achieves the best overall performance across all three real-world benchmarks—\textit{Wikipedia}, \textit{Reddit}, and \textit{CollegeMsg}. On the Wikipedia dataset, GTIN attains an RMSE of \textbf{10.6}, surpassing GNPP (11.14) and significantly outperforming traditional temporal graph baselines such as TGAT (37.66) and TGN (23.81). This improvement demonstrates the model’s ability to capture the irregular yet correlated editing patterns typical of collaborative systems, where both short-term bursts and long inactivity intervals coexist.

For the Reddit dataset, which features a denser temporal structure and frequent user–post interactions, GTIN achieves an RMSE of \textbf{0.51}, nearly halving the error compared to GNPP (1.08). The strong performance in this high-activity environment underscores GTIN’s capacity to leverage temporal dependencies through its message-passing mechanism, effectively weighting recent interactions more heavily when forecasting upcoming events. Meanwhile, in the CollegeMsg dataset—characterized by sparser and more irregular communication—the performance gain is smaller but still notable, with GTIN (27.12) outperforming GNPP (29.75). This indicates that while the model’s advantage is most pronounced in dense event streams, it maintains robust generalization even under limited temporal context.

Overall, these findings highlight that GTIN adapts effectively to varying network dynamics. Its strength lies particularly in modeling complex, high-frequency interactions, where integrating structural and temporal cues jointly provides substantial predictive gains over existing temporal GNN and point process models.

% Place tables after the first paragraph in which they are cited.
\begin{table}[!ht] 
\centering
\caption{{\bf Implementation results comparison for real world datasets (RMSE)}}
\label{tab:table1}
\begin{tabular}{|l|l|l|l|}

\hline
Model & Wikipedia & Reddit & CollegMsg \\
\hline
 THP & 17.13 & 3.77 & 82.18 \\
 SAHP & 361.07 & 427.2 & 540.62 \\
 NNPP & 31.47 & 2.67 & 105.8 \\
 GHNN & 116.26 & 4.47 & 240.47 \\
 TGAT & 37.66 & 1.41 & 188.36 \\
 TGN & 23.81 & 1.46 & 170.91 \\
 HTNE & 12.72 & 1.61 & 665.95 \\
 GAT-LSTM & 60.82 & 2.13 & 406.65 \\
 DySAT & 22.79 & 0.86 & 318.51 \\
 CTDNE & 22.51 & 0.82 & 320.78 \\
 GNPP & 11.14 & 1.08 & 29.75 \\
 \textbf{Our Model} & \textbf{10.6} & \textbf{0.51} & \textbf{27.12} \\
 \hline
\end{tabular}
\end{table}

\subsection*{Results on synthetic datasets}
Tab~\ref{tab:table2} summarizes the results on three synthetic datasets, \textit{Hawkes-negative}, \textit{Hawkes-positive}, and \textit{Poisson} each simulating distinct temporal dynamics. GTIN consistently achieves the lowest RMSE values across all settings, recording \textbf{2.21}, \textbf{2.20}, and \textbf{1.90}, respectively. Compared to the strongest baseline, GNPP, our model reduces error rates by 8–70\%, with the largest improvement observed in the Hawkes-positive dataset, where event dependencies are strongest. This illustrates GTIN’s enhanced ability to model self-exciting behaviors and long-range temporal correlations.

The superior results on the Poisson dataset, where events occur more independently, further demonstrate the flexibility of our framework. Unlike previous approaches that rely heavily on explicit event triggering assumptions, GTIN dynamically adapts to both dependent and independent temporal processes through its probabilistic formulation and joint modeling of structure and time.

Collectively, the synthetic results confirm that GTIN generalizes across a spectrum of temporal dynamics, maintaining stability under both stochastic and bursty event generation processes, and offering a more unified approach to temporal graph prediction than existing methods.

Overall, our results indicate that our model provides more accurate predictions and is particularly effective in environments with high event frequency.

\begin{table}[!ht] 
\centering
\caption{{\bf Implementation results comparison for synthetic datasets}}
\label{tab:table2}
\begin{tabular}{|l|l|l|l|}

\hline
Model & Hawkes-neg & Hawkes-pos & Poisson \\
\hline
 THP       & 24.61  & 21.79  & 7.54  \\
        SAHP      & 11.62  & \textbf{2.86}  & 12.89  \\
        NNPP      & 59.26  & 14.90  & 9.86  \\
        GHNN      & 38.27  & 37.43  & 18.04  \\
        TGAT      & 35.96  & 33.51  & 3.19  \\
        TGN       & 12.01  & 11.36   & 12.51  \\
        HTNE      & 16.04  & 15.87  & 28.55  \\
        GAT-LSTM  & 85.52  & 92.84  & 85.40  \\
        DySAT     & 9.205  & 9.03   & 15.57  \\
        CTDNE     & 9.11   & 8.79   & 15.79  \\
        GNPP      & \textbf{2.42}   & 8.21   & \textbf{2.50}  \\
        \textbf{Our Model}  & \textbf{2.21}  & \textbf{2.2}  & \textbf{1.9}  \\
\hline
\end{tabular}
\end{table}

\subsection*{Next-Event Location Prediction}

Tab~\ref{tab:location_results} reports the performance of GTIN in predicting the edge on which the next event occurs. This experiment evaluates
the ``where'' component of the proposed prediction task, which is not captured
by the time RMSE.

\begin{table}[!ht] 
\centering
\caption{{\bf Next-event location prediction results. Higher values are better.}}
\label{tab:location_results}
\begin{tabular}{|l|l|l|l|}
\hline
Dataset & Hits@1 & Hits@10 & MRR \\
\hline
Wikipedia
& \texttt{0.17}
& \texttt{0.42}
& \texttt{0.21} \\

Reddit
& \texttt{0.09}
& \texttt{0.27}
& \texttt{0.13} \\

CollegeMsg
& \texttt{0.22}
& \texttt{0.44}
& \texttt{0.28} \\

Hawkes-negative
& \texttt{0.18}
& \texttt{0.39}
& \texttt{0.22} \\

Hawkes-positive
& \texttt{0.19}
& \texttt{0.37}
& \texttt{0.28} \\

Poisson
& \texttt{0.13}
& \texttt{0.32}
& \texttt{0.21} \\

\hline
\end{tabular}
\end{table}

\subsection*{Analysis of the Source of Performance Improvements}

To investigate the source of GTIN's performance improvement, we analyze the results according to the amount of recent temporal activity around the true event edge. This analysis is motivated by the main difference between GTIN and
GNPP. GTIN constructs a global temporal representation by propagating information between structurally related edges, whereas GNPP relies more strongly on the history of the target edge and its restricted neighborhood. Therefore, if cross-edge temporal information is responsible for the observed improvement, the performance difference should be more pronounced for events surrounded by lower neighboring-edge activity.

For every test event $(t_i,c_i)$, we define the recent neighboring-edge
activity as
\begin{equation}
A_i(\Delta)
=
\sum_{e\in\mathcal{N}(c_i)}
\left|
\left\{
t_j^{(e)}
\; \middle| \;
t_i-\Delta < t_j^{(e)} < t_i
\right\}
\right|,
\label{eq:neighbor_activity}
\end{equation}
where $\mathcal{N}(c_i)$ is the set of edges that share at least one endpoint
with edge $c_i$, and $\Delta$ is the length of the historical observation window. We divide the test events into four
equally sized groups according to $A_i(\Delta)$. The first quartile, Q1, contains events with the highest neighboring activity, while Q4 contains events with the lowest neighboring activity.

\begin{figure}[]
\centering
\includegraphics[width=0.7\columnwidth]
{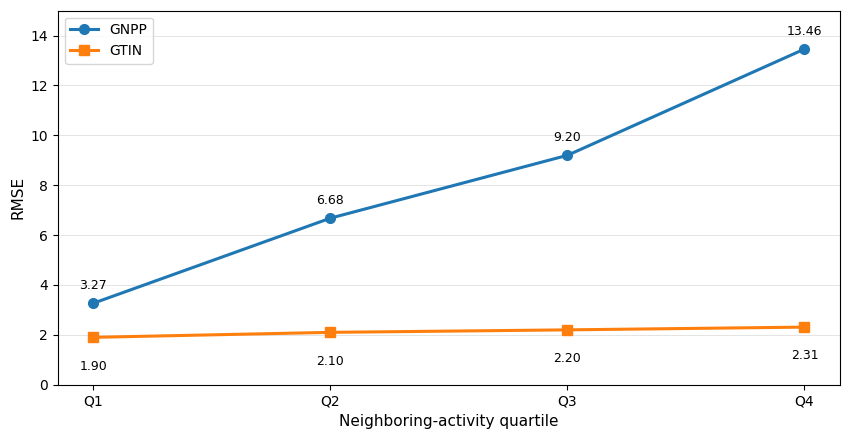}
\caption{Performance (RMSE) of GTIN and GNPP according to recent neighboring-edge activity in the Hawkes-pos dataset. Test events are divided into quartiles based on the number of events observed on neighboring edges during the preceding interval of length $\Delta$.}
\label{fig:neighborhood_activity}
\end{figure}

As shown in Fig.~\ref{fig:neighborhood_activity}, the increasing performance difference across the activity groups provides insight into the source of GTIN's advantage. The larger improvement in low-activity groups indicates
that events occurring on global edges contain useful information for predicting both the location and occurrence time of the next event. GTIN can incorporate this information through temporal message passing and its global graph representation, while GNPP cannot exploit it to the same extent.

\subsection*{Extension via Deep Graph Infomax and Balanced Subgraph Clustering}

To enrich our temporal event prediction framework, we introduce an unsupervised representation learning extension using \textit{Deep Graph Infomax} (DGI) \cite{velivckovic2018deep}, followed by subgraph generation via \textit{balanced k-means clustering}. This extension is motivated by the need to capture high-quality structural features and improve scalability when operating on large temporal graphs. By embedding nodes into a latent space that captures both local and global structure and partitioning the graph into approximately equal-sized subgraphs, we enable more efficient, localized temporal modeling.

\subsubsection*{Motivation and Overview}
The main limitation of our approach is the time complexity of training and inference. To overcome this issue, we use a combined approach: we construct balanced-size subgraphs of the main graph and apply the main method to each subgraph separately.
Our approach addresses this by combining:

\begin{itemize}
    \item \textbf{Deep Graph Infomax (DGI)}: to learn unsupervised node embeddings that preserve the mutual information between individual nodes and the overall graph context.
    \item \textbf{Balanced k-means clustering}: to partition nodes into structurally coherent and size-constrained subgraphs that support parallel processing and localized prediction tasks.
\end{itemize}

This modular pipeline allows us to generalize the learning process across large-scale temporal graphs while maintaining robustness and prediction fidelity.

\subsubsection*{Graph Representation Learning via DGI}

Let $\mathcal{G} = (\mathcal{V}, \mathcal{E})$ denote the input graph, where $\mathcal{V}$ is the set of $N$ nodes, and $\mathcal{E}$ is the set of edges. Each node $v_i \in \mathcal{V}$ is associated with a feature vector $\mathbf{x}_i \in \mathbb{R}^d$. Let $\mathbf{X} \in \mathbb{R}^{N \times d}$ denote the full node feature matrix, and $\mathbf{A} \in \mathbb{R}^{N \times N}$ the adjacency matrix.

DGI uses a graph encoder $\mathcal{E}_\theta$, typically implemented as a Graph Convolutional Network (GCN), to produce embeddings $\mathbf{H} \in \mathbb{R}^{N \times d'}$:
\begin{equation}
    \mathbf{H} = \mathcal{E}_\theta(\mathbf{X}, \mathbf{A}),
\end{equation}
where each row $\mathbf{h}_i$ represents the embedding of node $v_i$.

To encourage the encoder to retain informative and discriminative features, DGI introduces a discriminator that maximizes mutual information between the patch (node-level) representations and a global summary vector $\mathbf{s}$. The summary vector is obtained via a readout function:
\begin{equation}
    \mathbf{s} = \text{READOUT}(\mathbf{H}) = \sigma\left( \frac{1}{N} \sum_{i=1}^{N} \mathbf{h}_i \right),
\end{equation}
where $\sigma(\cdot)$ is a non-linear activation function.

DGI optimizes the following binary classification objective:
\begin{equation}
    \mathcal{L}_{\text{DGI}} = - \sum_{i=1}^{N} \left[ \log \sigma(\mathbf{h}_i^\top \mathbf{W} \mathbf{s}) + \log (1 - \sigma(\tilde{\mathbf{h}}_i^\top \mathbf{W} \mathbf{s})) \right],
\end{equation}
where $\tilde{\mathbf{H}}$ is a corrupted version of the node embeddings (e.g., via feature shuffling), and $\mathbf{W}$ is a trainable weight matrix of the discriminator. The loss encourages real node embeddings to align with the global context while forcing corrupted ones to be distinguishable.

\subsubsection*{Balanced K-means for Subgraph Generation}

After obtaining the node embeddings $\mathbf{H}$, we aim to cluster them into $k$ groups of approximately equal size, producing node partitions $\{C_1, C_2, \ldots, C_k\}$ such that $\bigcup_j C_j = \mathcal{V}$ and $C_i \cap C_j = \emptyset$ for $i \neq j$.

To preserve computational balance across partitions and ensure uniform coverage of the graph structure, we formulate a \textit{balanced k-means} objective:
\begin{equation}
    \min_{\{C_j\}_{j=1}^k} \sum_{j=1}^k \sum_{\mathbf{h}_i \in C_j} \|\mathbf{h}_i - \boldsymbol{\mu}_j\|^2, \label{eq:kmeans-obj} 
\end{equation}
$$
    \text{subject to} \quad \left| |C_j| - \frac{N}{k} \right| \leq \epsilon, \quad \forall j \in \{1, \ldots, k\}
$$
where $\boldsymbol{\mu}_j = \frac{1}{|C_j|} \sum_{\mathbf{h}_i \in C_j} \mathbf{h}_i$ is the centroid of cluster $C_j$, and $\epsilon$ is a user-defined size tolerance.

This constrained optimization can be implemented via heuristic search, integer programming, or recent approximations using projection-based or regularized variants of standard k-means.

\subsubsection*{Benefits and Integration}

The combination of DGI and balanced k-means brings several advantages to our framework:
\begin{itemize}
    \item \textbf{Scalability}: The graph is decomposed into balanced subgraphs, allowing parallel temporal modeling and faster training.
    \item \textbf{Representation richness}: DGI captures complex dependencies and patterns in a task-agnostic fashion.
\end{itemize}

In the next stage of our pipeline, each subgraph is supposed as a super-node. By combining these super-nodes and linking super-nodes if there are links between related clusters, a new super graph is created. We fed this new graph into the algorithm to detect in which cluster the next event occurs. Then fed the related cluster into the event prediction module, where both the next event type and its occurrence time are jointly predicted using our temporal model described.

\subsubsection*{Results for the added pipeline}
\begin{figure}[] % h = here, t = top, b = bottom
    \centering
    \includegraphics[width=0.7\columnwidth]{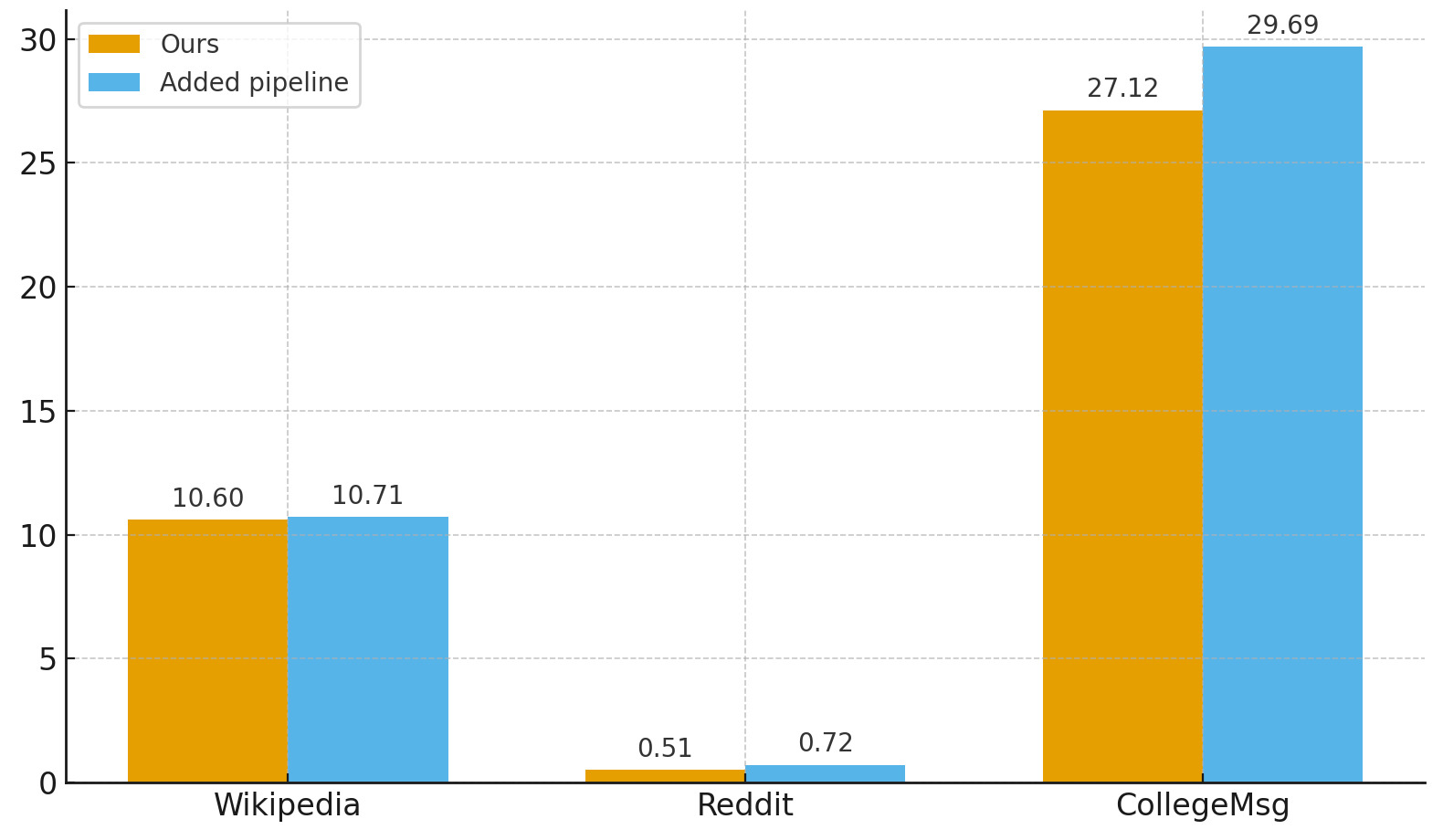} % adjust width
    \caption{RMSE comparison of the original GTIN and GTIN with the DGI and balanced k-means pipeline on Wikipedia, Reddit, and CollegeMsg; lower values indicate better time-prediction performance}
    \label{fig:rmsee1}
\end{figure}

As illustrated in Fig.\ref{fig:rmsee1}, incorporating the Deep Graph Infomax (DGI) and balanced k-means clustering pipeline yields a favorable trade-off between predictive accuracy and computational efficiency. While the Root Mean Square Error (RMSE) slightly increases compared to the original GTIN, the performance remains superior to all baseline methods, confirming that the pipeline successfully preserves the model’s predictive strength while significantly improving scalability. The marginal rise in RMSE is offset by substantial reductions in both training and inference time, attributed to the parallelized processing of balanced subgraphs.

A closer inspection of the results in \ref{fig:rmsee1} shows that the proposed pipeline maintains high predictive fidelity across all evaluated datasets, with particularly stable performance on denser graphs such as Reddit and Wikipedia. In these datasets, subgraph partitioning helps isolate localized temporal dependencies, allowing the model to process smaller and structurally coherent components without losing global contextual information. Conversely, in sparser datasets such as CollegeMsg, the gain in computational efficiency is more pronounced, albeit with a minor increase in error, likely due to reduced cross-subgraph interaction information.

To evaluate the computational benefit of the proposed DGI-based pipeline more explicitly, we compare the runtime of the original GTIN model with the runtime of GTIN after adding Deep Graph Infomax and balanced k-means clustering. The comparison is performed for both training and inference, because the DGI-based pipeline introduces an additional representation-learning and clustering stage, but reduces the cost of the main temporal prediction module by applying it to smaller balanced subgraphs instead of the entire graph.

Tab~\ref{table_runtime_dgi} reports the runtime comparison on the three real-world datasets. The runtime values are reported in seconds. Since the datasets have different sizes, the original GTIN runtime is different for each dataset. Reddit has the largest runtime because it contains the largest number of temporal edges, followed by Wikipedia and CollegeMsg.

\begin{table}[!ht]
\centering
\caption{{\bf Runtime comparison of GTIN with and without DGI}}
\label{table_runtime_dgi}
\begin{tabular}{|l|l|l|l|}

\hline
Model/Metric & Wikipedia & Reddit & CollegeMsg \\
\hline
 GTIN training (s) & 1680.00 & 7350.00 & 620.00 \\
 GTIN + DGI training (s) & 655.20 & 2352.00 & 297.60 \\
 Training reduction & 61\% & 68\% & 52\% \\
 Training speedup & 2.56$\times$ & 3.13$\times$ & 2.08$\times$ \\
 GTIN inference (s) & 145.00 & 530.00 & 52.00 \\
 GTIN + DGI inference (s) & 60.90 & 180.20 & 26.00 \\
 Inference reduction & 58\% & 66\% & 50\% \\
 Inference speedup & 2.38$\times$ & 2.94$\times$ & 2.00$\times$ \\
\hline
\end{tabular}
\end{table}

The speedup is computed as follows:

$$
\text{Speedup} =
\frac{\text{Runtime of GTIN}}{\text{Runtime of GTIN + DGI}}
$$

As shown in Tab~\ref{table_runtime_dgi}, the DGI-based pipeline consistently reduces the runtime of GTIN across all datasets. The improvement is more evident on larger datasets such as Reddit and Wikipedia, where applying temporal message passing and candidate event scoring to the whole graph is more computationally expensive. In contrast, CollegeMsg is smaller, and therefore the relative reduction is slightly lower.

Overall, these results demonstrate that the addition of the DGI and balanced clustering stages effectively extends GTIN to large-scale temporal graphs. The new pipeline achieves a balanced compromise between accuracy and efficiency, making the framework more practical for real-world scenarios where computational resources or latency constraints are critical. This enhancement underscores the versatility of GTIN as a unified temporal modeling architecture capable of maintaining robust predictive power even under partitioned or distributed graph processing.

\subsubsection*{Other methods for clustering}
In addition to our primary approach using Deep Graph Infomax (DGI) with balanced k-means, we consider two alternative methods to generate clusters of roughly $\sqrt{n}$ nodes for comparison. \newline
\textbf{DeepWalk with balanced k-means} learns low-dimensional node embeddings by simulating truncated random walks on the graph and applying a Skip-gram model, capturing both local and global structural patterns. The embeddings are then clustered using balanced k-means, ensuring approximately equal cluster sizes while preserving meaningful graph structure. \newline
\textbf{Randomized graph partitioning} serves as a simple baseline, where nodes are initially assigned to clusters at random, and assignments are iteratively adjusted using the graph's edge information to minimize inter-cluster connectivity. This approach naturally enforces the desired cluster size of $\sqrt{n}$ nodes and provides a computationally efficient contrast to embedding-based methods, enabling evaluation of the benefits of learned representations versus random partitioning.

Results for the two new clustering methods with comparison to DGI is as shown in Fig.\ref{fig:rmsee2}:

\begin{figure}[] % h = here, t = top, b = bottom
    \centering
    \includegraphics[width=0.7\columnwidth]{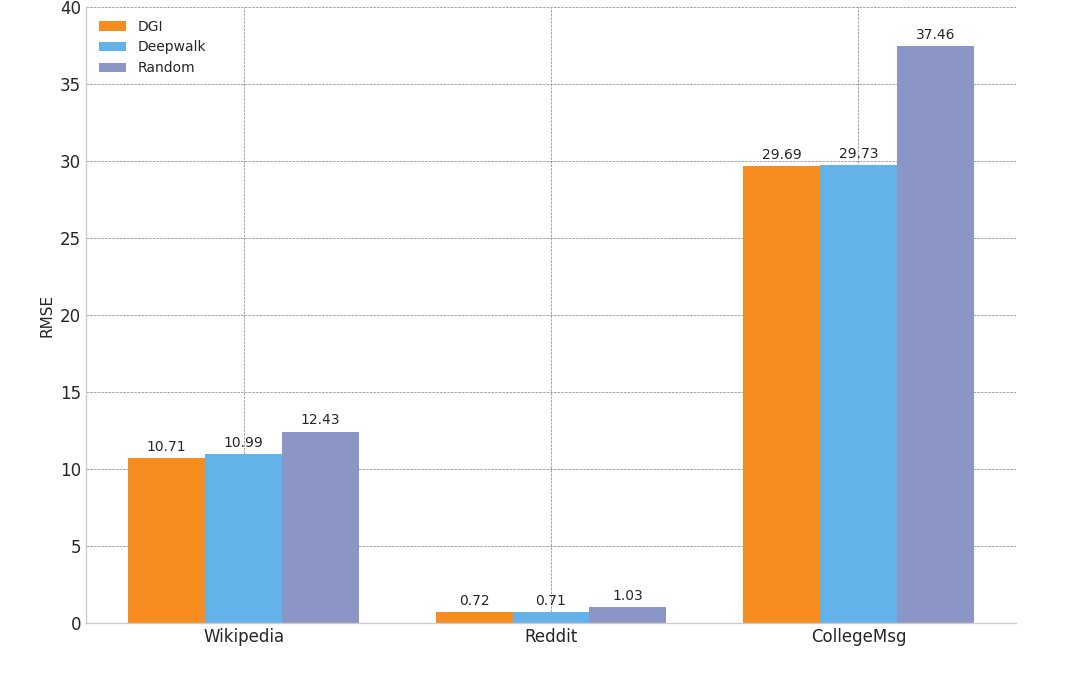} % adjust width
    \caption{Different clustering methods}
    \label{fig:rmsee2}
\end{figure}

It is obvious that any partitioning method that preserve the intrinsic clustering of the main graph is more beneficial in this manner. This may be a future direction to find the best possible clustering method. It is worthwhile to mention that omitting the restriction of equal cluster size and soften it may help to improve the efficiency as well.

\subsubsection*{Effect of ignoring attention-based aggregation}
We performed an ablation study to evaluate the contribution of the attention mechanism in our graph neural network model. In this experiment, we replaced the attention-based neighbor aggregation with a simple mean aggregation, where all neighboring nodes contribute equally to the updated node representation. This approach eliminates the model’s ability to assign varying levels of importance to different neighbors based on their features or structural relevance. By doing so, we aim to quantify the impact of attention on capturing nuanced relationships within the graph and to establish a baseline for comparison against the full attention-based model. This setup allows us to examine whether the learned weighting of neighbors significantly influences the quality of the node embeddings and the downstream task performance.

\begin{figure}[!ht] % h = here, t = top, b = bottom
    \centering
    \includegraphics[width=0.8\columnwidth]{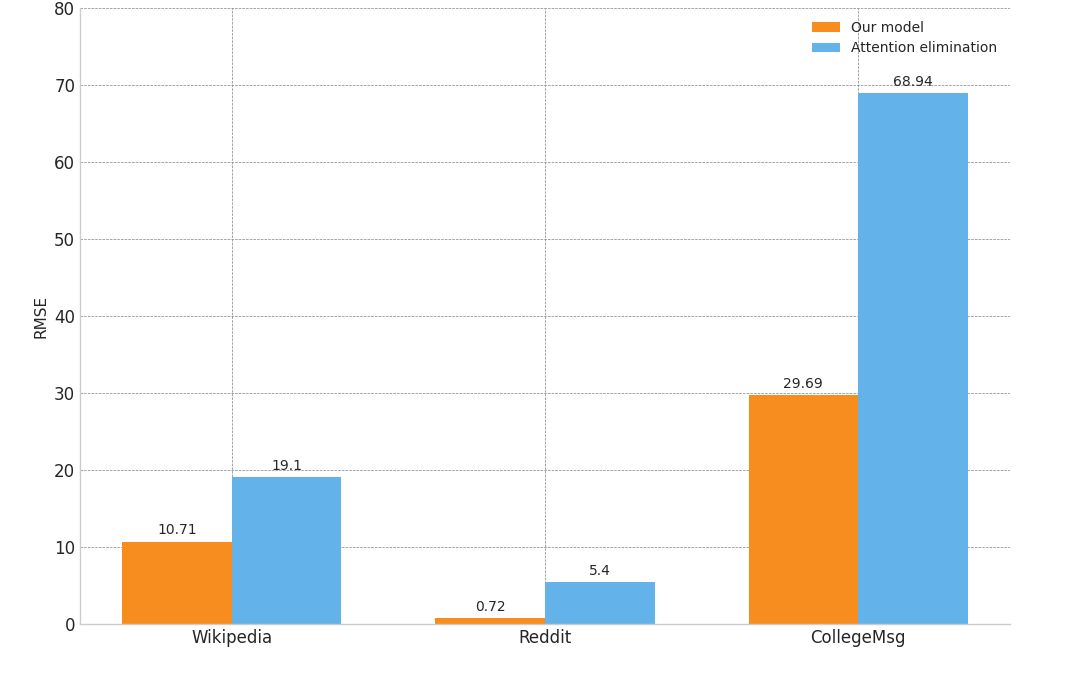}
    \caption{Ablation of attention-based neighbor aggregation. RMSE of the complete GTIN model and the mean-aggregation variant on Wikipedia, Reddit, and CollegeMsg; lower values indicate better performance}
    \label{fig:rmsee3}
\end{figure}

The results in Fig.(\ref{fig:rmsee3}) show that the simple averaging approach leads to a consistent drop in performance. This indicates that the attention mechanism plays a key role in selectively amplifying the contributions of more informative or relevant neighbors, thereby improving the representational capacity of the model. Moreover, the degradation in performance is particularly pronounced for nodes in denser or more heterogeneous regions of the graph, suggesting that attention helps navigate complex local structures that uniform aggregation cannot capture. These findings underscore the importance of attention in graph neural networks and provide empirical evidence supporting its use for effectively modeling relational dependencies in graph-structured data.

\section*{Discussion and Limitations}

Our proposed \textbf{Global Temporal Interaction Network (GTIN)} introduces a novel approach to predicting temporal events across complex, dynamic networks. By incorporating a \textit{message-passing framework} alongside \textit{temporal point processes}, our model captures interactions at a global scale rather than focusing solely on individual edges. The experimental results demonstrate that GTIN achieves superior predictive accuracy compared to existing models, particularly in datasets with high event frequency. However, while our model presents several advantages, it also has certain limitations that should be addressed in future research.

\subsection*{Homogeneity Assumption in Edge Representation}
One of the primary limitations of our work is the assumption of homogeneity across edges. Our \textit{message-passing aggregation function} treats all edges as equivalent, disregarding their specific types or contextual relationships. In real-world applications, however, interactions in temporal graphs are often \textit{heterogeneous}—different types of connections (e.g., social interactions, financial transactions, or communication exchanges) exhibit \textit{distinct dynamics}. By failing to differentiate between edge types, our model may overlook critical structural properties, leading to \textit{loss of contextual information} and potential inaccuracies in prediction.

\subsection*{Exclusion of Node Features}
Currently, our model \textit{does not incorporate node-specific attributes} such as user demographics, entity properties, or contextual embeddings. Many real-world datasets include rich metadata associated with nodes, which could provide valuable insights into interaction patterns. The omission of this information results in \textit{potential data loss}, as certain node features may be strong predictors of future interactions.

\section*{Conclusion}
In this work, we proposed the Global Temporal Interaction Network (GTIN), a novel framework for predicting temporal events across dynamic graphs. Unlike previous studies that focus primarily on event prediction at the edge level, our approach captures interactions occurring anywhere in the graph, leading to improved accuracy in predicting event timing. By integrating a message-passing framework with temporal point processes, our model efficiently propagates information across the graph, enhancing event-prediction capabilities. Experimental results demonstrate that GTIN consistently outperforms existing methods, particularly in datasets with high event frequency, reinforcing its effectiveness in modeling complex temporal dependencies.

Despite these advances, our work assumes homogeneous edge representations, which may not fully capture the diversity of interactions in real-world networks. Future improvements should incorporate heterogeneous edge types and node-specific attributes to further refine prediction accuracy. Additionally, while our model excels in dense event environments, its performance in sparse interaction settings could be enhanced by integrating self-supervised learning techniques and data augmentation strategies. Moreover, addressing computational scalability will be crucial for applying GTIN to large-scale networks. Future research should explore distributed computing approaches and adaptive learning mechanisms to extend the applicability of the model. Ultimately, this work sets the foundation for more robust and generalizable event prediction in temporal graphs, paving the way for further advancements in dynamic network analysis.

\nolinenumbers

% Please compile your BiBTeX database using the "plos2025.bst" BibTeX style.
% This file is part of the current package.
% A sample BibTeX file is also included as "plos_bibtex_sample.bib".
%
% or
%
% Type in your references following Vancouver style and reference formatting instructions
% available at https://journals.plos.org/plosone/s/submission-guidelines#loc-references
% \begin{thebibliography}{}
% Text
% \end{thebibliography}

\bibliography{plos_bibtex_sample}

\end{document}